\documentclass[12pt]{article}

\usepackage{amssymb, amsmath}
\usepackage{graphicx}
\usepackage{cite}

\newcommand{\td}{\text{d}}

\def\be{\begin{equation}}
\def\ee{\end{equation}}
\def\bea{\begin{eqnarray}}
\def\eea{\end{eqnarray}}

\title{ \bf{Black hole non-uniqueness via spacetime topology in five dimensions}}

\author{Hari K. Kunduri$^a$\footnote{hkkunduri@mun.ca } \  and James Lucietti$^b$\footnote{j.lucietti@ed.ac.uk } \\ \\
\small \sl $^a$ Department of Mathematics and Statistics, \\  \small \sl Memorial University of Newfoundland \\ \small \sl St John's NL A1C 4P5, Canada
\\ \small \sl $^b$  School of Mathematics and Maxwell Institute of Mathematical Sciences, \\ \small \sl University of Edinburgh, \\ \small \sl   King's Buildings, Edinburgh, EH9 3JZ, UK }

\date{}

\begin{document}

\maketitle


\begin{abstract}
The domain of outer communication of five-dimensional asymptotically flat stationary spacetimes
may possess non-trivial 2-cycles. We discuss how this may lead to a gross violation of black hole uniqueness, beyond the existence of black rings, even for solutions with two commuting rotational symmetries.  We illustrate this with a simple example in minimal supergravity; a four parameter family of supersymmetric black hole solutions, with spherical horizon topology and a 2-cycle in the exterior. We show there are black holes in this family with identical conserved changes to the BMPV black hole, thereby demonstrating black hole non-uniqueness in this context.  We find a decoupling limit of this family of black holes that yields spacetimes asymptotic to the near-horizon geometry of a BMPV black hole which contain a black hole and an exterior 2-cycle.
\end{abstract}

\newpage


\section{Topology of black hole spacetimes}
\label{sec:topology}

Topological censorship implies that the domain of outer communication of any globally hyperbolic asymptotically flat spacetime is simply connected~\cite{Topcen}. In four dimensions this is sufficient to guarantee that any spatial hypersurface $\Sigma$ has the trivial topology $\mathbb{R}^3 \backslash B$, where $B$ is the black hole region. This can then be used to show that $\partial B$ must be homeomorphic to a sphere $S^2$, an important ingredient of the black hole uniqueness theorem~\cite{Chrusciel:1994tr}.

For higher dimensional spacetimes, topological censorship is not as restrictive. In particular, spatial hypersurfaces $\Sigma$ may have non-trivial higher homology groups $H_p(\Sigma)$, for $ p\geq 2$. This raises the question: are there black hole spacetimes whose domain of outer communication  is topologically non-trivial? 

For static black holes this cannot be the case; there is a uniqueness theorem~\cite{Gibbons:2002bh, Gibbons:2002av} that generalises the four dimensional case which shows the only solution is the Schwarzschild metric, or charged versions thereof, so $\Sigma \cong \mathbb{R}^{D-1} \backslash B$. Therefore, if there are any solutions with $\Sigma$ non-trivial they must be not be static.

In this note we will focus on five-dimensional asymptotically flat stationary spacetimes with two commuting rotational Killing fields, containing a single black hole.  In this case it has been shown that the topology of the domain of outer communication is $\mathbb{R}\times \Sigma$, where\footnote{In fact, the statement regarding $\Sigma$ is still true if only one rotational Killing field is assumed, although then there are more possibilities for the horizon topology~\cite{Hollands:2010qy}.} 
\be
\Sigma \cong \Big( \mathbb{R}^4 \# n (S^2\times S^2) \# n' (\pm \mathbb{CP}^2) \Big) \backslash B, \label{Sigma}
\ee  
for some $n,n' \in \mathbb{N}$ and $B$ is the black hole region, where the horizon $H\cong \partial B$ must be one of $S^3, S^1 \times S^2, L(p,q)$~\cite{Hollands:2007aj, Hollands:2008fm, Hollands:2010qy, Hollands:2012xy}. The integers $n,n'$ determine the 2-cycle structure of $\Sigma$. 

The known solutions consist of the Myers-Perry black holes and the black rings, which have horizon topologies $H \cong S^3$ and $H \cong S^1 \times S^2$ respectively, see~\cite{Emparan:2008eg} for a review. Both of these solutions have trivial topology in the exterior, i.e. $n=n'=0$ (see~\cite{Alaee:2013oja} for more detail on their topology). Thus, a more refined question is: are there black hole spacetimes with a domain of outer communication given by (\ref{Sigma}) for some $n, n' \neq 0$?

The above class of spacetimes belong to the generalised Weyl solutions~\cite{Emparan:2001wk}. These have been well studied and their classification is understood in terms of a so-called `rod structure'~\cite{Emparan:2001wk, Harmark:2004rm, Hollands:2007aj}. The rod structure is equivalent to the specification of the orbit space $\hat{\Sigma}\cong \Sigma \slash U(1)^2$, which turns out to be a manifold with boundaries and corners, together with a pair of integers and a real number for each boundary segment. These numbers correspond to the linear combination of Killing fields which vanishes along a given boundary segment and the length of the boundary segment. 

In particular, it has been shown that vacuum black hole spacetimes in this class are uniquely specified by their angular momenta and their rod structure (their mass is determined by these)~\cite{Hollands:2007aj, Hollands:2008fm, Figueras:2009ci}. The known solutions mentioned above possess the simplest possible rod structure compatible with their horizon topology. However, these results do not address the general existence problem: that is, for what rod structures do black hole solutions actually exist? In fact, this is a further refinement of the question posed above. This is because the orbit space and the integers for each boundary segment determine the manifold $\Sigma$ together with the $U(1)^2$-action, which in turn gives the integers $n,n'$ in the decomposition (\ref{Sigma}).

 An analogous uniqueness result for Einstein-Maxwell theory is known, under some restrictive assumptions, which reveals one must also specify the magnetic flux through any 2-cycles to uniquely characterize  the solution~\cite{Hollands:2007qf}. Various uniqueness results are also known in minimal supergravity (Einstein-Maxwell with a Chern-Simons term) for spherical black holes~\cite{Tomizawa:2009ua}, for black rings~\cite{Tomizawa:2009tb} and for multiple black holes~\cite{Armas:2009dd}. However, all of these assume the simplest possible rod structure --  and hence exterior topology -- which is compatible with the assumed horizon topology.

It is clear that the existence of black hole spacetimes containing non-trivial 2-cycles, often termed `bubbles', would represent a gross violation of black hole uniqueness beyond black rings. We emphasise this is even compatible with the $\mathbb{R}\times U(1)^2$ symmetry possessed by the known explicit solutions. This possibility does not appear to have been discussed before.

For vacuum gravity, the only mechanism available to `support' the bubbles from collapsing is rotation. However, it is unclear whether new vacuum black holes of this kind actually exist. We emphasise though that there does not appear to be any convincing reason why not; indeed, it is tempting to conjecture their existence. For Einstein gravity coupled to a Maxwell field, a more obvious mechanism is available for supporting bubbles: {\it magnetic flux}. Indeed, it is much easier to envisage a vast set of black hole solutions with bubbles in the exterior supported by magnetic flux.

In the absence of black holes,  soliton spacetimes with bubbles supported by flux are well known to exist, with a number of supersymmetric (see the review \cite{Bena:2007kg}) and non-supersymmetric examples \cite{Bena:2009qv,Compere:2009iy, Bobev:2009kn}. These spacetimes carry positive energy.  The relationship between the mass of these spacetimes and their fluxes is expressed in a Smarr-type formula, as observed for BPS-solitons by Gibbons and Warner \cite{Gibbons:2013tqa}. Recently, it was shown that on-shell variations of the mass and magnetic fluxes for general soliton spacetimes are governed by a `first law' formula \cite{Kunduri:2013vka}.

Furthermore, one can derive a generalised mass and mass variation formula  for $\mathbb{R} \times U(1)^2$-invariant spacetimes containing a black hole with an arbitrary number of bubbles in the exterior region. Similarly to the soliton case it was found that on top of the usual terms for a black hole, extra terms due to the bubbles are present. However, in contrast, these bubble terms are most naturally expressed in terms of an `electric' flux charge. For Einstein-Maxwell theory, possibly with a Chern-Simons term, the mass formula is \cite{Kunduri:2013vka},
\be
 M = \frac{3 \kappa A_H}{16 \pi} + \frac{3}{2} {\Omega}_i  {J}_i + \Phi_H Q+  \frac{1}{2}\sum_{[C]}  {\cal Q}[C]  \Phi[C]+   \frac{1}{2}\sum_{[D]}{\cal Q}[D]   \Phi[D]  \label{BHmass}
 \ee
 and the first law of black hole mechanics is,
 \be
 \delta M = \frac{ \kappa \delta A_H}{8\pi} + \Omega_i \delta J_i + \Phi_H \delta Q + \sum_{[C]} {\cal Q}[C]  \delta  \Phi[C] +\sum_{[D]}  {\cal Q}[D]  \delta \Phi[D]    \; .  \label{BHmech}
\ee In the above $[C]$ is a basis for the second homology of $\Sigma$, $[D]$ are certain disc topology surfaces which extend from the horizon, $\Phi$ are magnetic potentials and  $\cal{Q}$ are certain `electric' fluxes defined on these surfaces. This shows that non-trivial spacetime topology plays an important role even in black hole thermodynamics, thus providing further motivation to study such objects beyond the question of black hole uniqueness.

The purpose of this note is to point out that asymptotically flat black solutions with non-trivial exterior topology {\it do} exist. We will illustrate this by writing down and analysing an explicit supersymmetric example.

\section{A supersymmetric example}

Large families of supersymmetric smooth solitons with bubbles have been constructed in the `fuzzball' literature~\cite{Bena:2005va}. Because of supersymmetry, it is a simple matter to `add' a black hole to such `bubbling' geometries. This has not really been emphasised in the literature, presumably because the focus of the fuzzball program has been on smooth geometries rather than black holes.

 We will discuss the simplest possibility of an asymptotically flat black hole solution to minimal supergravity with $S^3$ horizon topology possessing one exterior bubble.  We will first write down the solution and then discuss its salient features.
 
 \subsection{Solution}
Supersymmetric solutions of ungauged minimal supergravity are well understood: they take the form
\begin{equation}
\td s^2 = -f^2( \td t + \omega)^2 + f^{-1} \td s^2_M \; ,
\end{equation} 
where $V = \partial /\partial t$ is the supersymmetric Killing vector field and $\td s^2_M$ is a hyperK\"ahler base~\cite{Gauntlett:2002nw}. We choose the base $M$ to be a Gibbons-Hawking space
\begin{equation}
\td s^2_M = H^{-1} (\td \psi + \chi_i \td x^i)^2 + H \td x^i \td x^i \; ,
\end{equation} 
where $x^i, i=1,2,3$, are Cartesian coordinates on $\mathbb{R}^3$, the function $H$ is harmonic on $\mathbb{R}^3$ and $\chi$ is a 1-form on $\mathbb{R}^3$ satisfying $\star_3 \td  \chi = \td H$. 

As is well known~\cite{Gauntlett:2002nw}, such solutions are then specified by 4 harmonic functions $H,K,L,M$, in terms of which
\begin{equation} 
 f^{-1} = H^{-1} K^2 + L \; ,\qquad  \omega = \omega_\psi(\td \psi + \chi_i \td x^i) + \hat{\omega}_i \td x^i \; ,   \label{f}
 \ee
 where 
 \begin{eqnarray}
 \label{omega}
&& \omega_\psi = H^{-2} K^3 + \frac{3}{2}H^{-1}KL + M \; , \\  && \star_3 \td \hat\omega = H \td M - M \td H + \frac{3}{2}(K \td L - L \td K)
   \; .
 \end{eqnarray} 
 The Maxwell field is then
 \be
 \label{max}
 F = \frac{\sqrt{3}}{2} \td \left[ f( \td t + \omega) - K H^{-1} (\td \psi+ \chi_i \td x^i)  - \xi_i \td x^i\right]  \; ,
 \ee
where the 1-form $\xi$ satisfies  $\star_3 \td \xi = - \td K$.

Now we write the $\mathbb{R}^3$ in polar coordinates
\be
\td x^i \td x^i = \td r^2 + r^2 (\td\theta^2 + \sin^2\theta \td\phi^2)  \; .
\ee
Following~\cite{Bena:2005va}, consider the solution given by the harmonic functions
\begin{eqnarray}
H &=& \frac{1}{r}  - \frac{1}{r_1}+ \frac{1}{r_2}, \qquad  \qquad K = \frac{k_0}{r} + \frac{k_1}{r_1} + \frac{k_2}{r_2} ,\\
L &=& 1+ \frac{\ell_0}{r}  + \frac{\ell_1}{r_1} + \frac{\ell_2}{r_2}, \qquad  \qquad M = m + \frac{m_0}{r}  + \frac{m_1}{r_1}+ \frac{m_2}{r_2},
\end{eqnarray} where
\be
r_1 = \sqrt{r^2 + a_1^2 - 2ra_{1} \cos\theta}, \qquad \qquad r_2 = \sqrt{r^2 + a_2^2 - 2ra_{2} \cos\theta}   \label{centres}
\ee
are the distances from the origin to the `centres' ${\bf x}_1=(0,0,a_1)$ and ${\bf x}_2=(0,0,a_2)$ respectively. We assume $0 < a_1 < a_2$.  We will use a shift freedom in the harmonic functions to set $m_0 =0$ without any loss of generality~\cite{Bena:2005va}.  

The explicit form of the solution involves determining the 1-forms $\chi$ and $\hat{\omega}$ defined above. Integrating, we find 
 \be
 \chi =\left[  \cos \theta - \frac{r\cos \theta -a_1 }{r_1}+ \frac{r \cos\theta - a_2}{r_2}  \right] \td\phi \; ,
 \ee 
 where we have absorbed any integration constants by suitably shifting $\psi$.  Finally, integrating \eqref{omega} we get
\be
\hat\omega  = \hat{\omega}_\phi \td\phi,
\ee 
where, 
\begin{eqnarray}
\hat{\omega}_\phi &=& \frac{1}{2r a_1}\left[(m_1+\tfrac{3}{2}(  \ell_1 k_0 -\ell_0 k_1)) \left(r_1+ \frac{(r^2-a_1^2)}{r_1} \right)-  (m-\tfrac{3}{2} k_1 )r \left( r_1  - \frac{(r^2-a_1^2)}{r_1} \right)\right]  \nonumber \\
&+&\frac{1}{2 r a_2}\left[ (m_2+\tfrac{3}{2}( \ell_2 k_0 -\ell_0 k_2))\left(r_2+\frac{(r^2-a_2^2)}{r_2} \right) +(m  + \tfrac{3}{2}k_2)r \left( r_2 - \frac{(r^2-a_2^2)}{r_2}\right) \right]     \nonumber \\
&-&\frac{(m_1 + m_2+\tfrac{3}{2}(\ell_1 k_2 - \ell_2 k_1))}{r_1r_2(a_2 - a_1)}\left[a_1 a_2 - (a_1 + a_2)r \cos\theta +r^2 \right]  - (m +\tfrac{3}{2} k_0)\cos\theta +c  \nonumber
\end{eqnarray} 
and $c$ is an integration constant.
It is also  worth noting that the 1-form $\xi$ which determines the Maxwell field is
\be
\xi  = -\left[  k_0 \cos \theta + \frac{k_1(r\cos \theta -a_1) }{r_1}+ \frac{k_2(r \cos\theta - a_2)}{r_2}  + c' \right] \td\phi  \; ,
 \ee 
 where $c'$ is a constant. 
 
 For a suitable choice of constants this solution is asymptotically flat. 
 Defining $r = \rho^2/4$, it is easy to check the Gibbons-Hawking base  for $\rho \to \infty$ looks like
 \bea
 &&\td s^2_M = \frac{\rho^2}{4} \left(1 + {\cal O}(\rho^{-2}) \right) [\td\psi+ (\cos \theta +\mathcal{O}(\rho^{-2}) ) \td\phi ]^2  \nonumber \\ && \qquad \;+ \left(1 + \mathcal{O}(\rho^{-2}) \right) \left( \td\rho^2 + \frac{\rho^2}{4}( \td \theta^2 + \sin^2 \theta \td\phi^2) \right)  \; ,
 \eea
which is therefore asymptotically $\mathbb{R}^4$ provided $\Delta \psi = 4\pi$, $\Delta \phi = 2\pi$ and $0 \leq \theta \leq \pi$. Now, it is also clear that $ f = 1+\mathcal{O}(\rho{^{-2}})$. Furthermore, it is easily verified that $\omega_\psi= \mathcal{O}(\rho^{-2})$ and $\hat{\omega}_\phi = \mathcal{O}
(\rho^{-2})$ provided one chooses the constants
\begin{equation}
 m = -\frac{3}{2}(k_0 + k_1 +k_2)   \label{m}
 \end{equation}
and 
\begin{eqnarray} 
\label{c2_AF}
 &&c = \frac{1}{2(a_1-a_2)} \left[  \frac{3 a_1}{a_2}(\ell_0 k_2 - \ell_2 k_0 - \frac{2}{3} m_2) - \frac{3a_2}{a_1}(\ell_0 k_1 - \ell_1 k_0 - \frac{2}{3} m_1)  \right. \nonumber \\ && \quad \qquad \qquad \qquad \left. + 3 \ell_0(k_1 - k_2) - 3\ell_1(k_0 + k_1) - 4m_1 \right] \; ,
 \end{eqnarray}
 respectively.  With these choices, which will be assumed henceforth, our five dimensional spacetime is asymptotically flat $\mathbb{R}^{1,4}$.

 The above solution to supergravity is simply the explicit form for a supersymmetric solution whose harmonic functions possess three `centres'. If the parameters of the solution are chosen so the three centres are smooth timelike points, the resulting solution is the known soliton with two bubbles~\cite{Bena:2005va}. Instead, we will choose one of the centres to correspond to a horizon, and the other two to be smooth points.  We note  the above 3-centred solution is the simplest possibility for a black hole with an exterior 2-cycle.
 
Before moving on, we observe that if we `remove' two of the centres by taking the limit $a_1 \to a_2$ and  setting $k_1=k_2=\ell_1=\ell_2=m_1=m_2=0$ the solution reduces to the 1-centred BMPV black hole~\cite{BMPV, Gauntlett:1998fz}.

 \subsection{Regularity and causality}\label{sec:regularity}
 
 Let us first examine regularity at the centres ${\bf x}_1$ and ${\bf x}_2$.  We wish to impose that the spacetime metric at these centres is timelike and smooth.

To examine the behaviour of the metric near each of the centres we first introduce spherical polar coordinates for the $\mathbb{R}^3$ base adapted to each centre: $r_i = |{\bf x} - {\bf x}_i|$, $\cos \theta_i = (z-a_i)/r_i$ and $\phi_i =(-1)^i \phi$ for $i=1,2$. Now define  $R_i= 2\sqrt{r_i}$. Then one finds the Gibbons-Hawking base
\bea
\td s^2_M &=& F_i \left[ \td R_i^2+ \tfrac{1}{4} R_i^2 \left( \td \theta_i^2+\sin^2\theta_i \td \phi_i^2 + \frac{1}{F_i^2} ( \td \psi+( \cos\theta_i+ G_i) \td \phi_i)^2\right) \right]  \; ,
\eea
where we have defined  $F_i \equiv \tfrac{1}{4}R_i^2 H$ and $G_i \equiv (-1)^i\chi_\phi - \cos \theta_i $. It is convenient to rewrite the base as
\bea
\td s^2_M &=& F_i \left[ \td R^2_i + \frac{1}{4} R^2_i\left( (\td\psi + \cos \theta_i \td\phi_i)^2 + \td \theta_i^2 +\sin^2 \theta_i \td \phi_i^2 \right) \right]   \label{R4base} \\  &+& \tfrac{1}{4} F_i \left[ R_i^2 \left(\frac{1}{F_i^2}-1 \right) ( \td \psi+ \cos\theta_i \td \phi_i)^2  + \frac{2R_i^2G_i}{F_i^2} ( \td \psi +\cos \theta_i \td\phi_i) \td \phi_i + \frac{R_i^2G_i^2}{F_i^2} \td \phi_i^2 \right]  \; .  \nonumber
\eea
Observe that the metric in the square brackets on the first line is $\mathbb{R}^4$ in spherical polar coordinates, provided we make the same identifications on the angles as required for asymptotic flatness discussed above. Also, it is easily verified that $F_i = (-1)^i+ \mathcal{O}(R_i^2)$ and $G_i = \mathcal{O}(R_i^4)$ as $R_i \to 0$, so that the terms in the square brackets on the second line are subleading. This shows that, up to an overall sign, the base near the centres looks like a regular origin of $\mathbb{R}^4$. Of course, we only require the spacetime metric to have the correct signature.

To examine smoothness near the centres it is useful to introduce the coordinates 
\be
X_i = R_i \cos (\tfrac{1}{2} \theta_i),  \qquad Y_i = R_i \sin (\tfrac{1}{2} \theta_i ),  \qquad \phi_i^\pm = \tfrac{1}{2}(\psi \pm \phi_i) \; ,
\ee
in terms of which
\be
\td s^2 (\mathbb{R}^4) = \td X_i^2 + X_i^2 (\td \phi_i^+)^2  + \td Y_i^2 + Y_i^2 (\td \phi_i^-)^2   \; . 
\ee
Observe that $\Delta \phi_i^\pm = 2\pi$ and $X_i, Y_i$ are radial variables in two orthogonal planes of $\mathbb{R}^4$, so any $U(1)^2-$invariant smooth function on $\mathbb{R}^4$ must be a smooth function of $X_i^2, Y_i^2$ alone.
Indeed, it can be checked that $F_i, G_i$ are analytic in $X_i^2$ and $Y_i^2$ and in particular $G_i = \mathcal{O}(X_i^2 Y_i^2)$.  In these coordinates, the terms in the square brackets on the second line of (\ref{R4base}) are,
\bea
&&\frac{4}{R_i^2}\left(\frac{1}{F_i^2}-1 \right) (X_i^2 \td \phi_i^+ + Y_i^2 \td \phi_i^- )^2   + \frac{1}{F_i^2} (G_i\td \phi_i^+ - G_i \td \phi_i^-)^2\nonumber \\ &&+ \frac{4}{F_i^2} (X_i^2 \td \phi_i^+ +  Y_i^2 \td \phi_i^- ) (G_i\td \phi_i^+ - G_i \td \phi_i^-) \nonumber \\
&&= \mathcal{O}(X_i^4) (\td \phi_i^+)^2 + \mathcal{O}(X_i^2 Y_i^2) \td \phi_i^+ \td \phi_i^- +\mathcal{O}(Y_i^4) (\td \phi_i^-)^2  \; ,
\eea
where to obtain the equality we have used the behaviour of $F_i, G_i$ near the centre. This is indeed smooth at the origin of $\mathbb{R}^4$ and thus shows that the base metric is smooth at each centre.
 
 Now, since $ V^2 = -f^2$,  demanding the centres to be timelike requires $f\neq 0$ at the centres, which corresponds to setting
 \be
 \ell_1 =  k_1^2, \qquad \qquad \ell_2 = -k_2^2  \; .  \label{ell12}
 \ee
 Then we see that in order to get the correct signature for the spacetime metric we will need $f|_{{\bf x}={\bf x}_2}>0$ and $f|_{{\bf x}={\bf x}_1}<0$.  Again, subleading terms of $f$ are analytic in $X_i^2, Y_i^2$, so the function $f$ is smooth at these centres.
 
 Enforcing smoothness of the metric near the centres also requires the invariant $V \cdot \partial_\psi = f^2 \omega_\psi$ to vanish at these points (since $\partial_\psi$ degenerates at these points). Firstly, requiring $\omega_\psi$ to be non-singular at the centres, and using (\ref{ell12}), implies
 \be
 m_1 = \frac{k_1^3}{2}, \qquad \qquad m_2 = \frac{k_2^3}{2}  \; .  \label{m12}
 \ee
 Then, requiring that $\omega_\psi$ actually vanishes as ${\bf x}\to {\bf x}_1$ and ${\bf x}\to {\bf x}_2$, implies 
 \be
\label{constr1}
 a_1(k_2 + k_1)^3 + (a_2 - a_1)(3k_0 k_1^2 +k_1^3  -3(k_0 + k_2 + 2k_1)a_1 -3k_1 \ell_0) = 0 \; , 
 \ee
 and
 \be
 \label{constr2}
 a_2(k_2 + k_1)^3 + (a_2 - a_1)(3k_0 k_2^2 - k_2^3 - 3(k_0 + k_1)a_2 + 3k_2 \ell_0) = 0  \; ,
\ee 
respectively. Note that these two constraints on the parameters have been simplified using (\ref{m}, \ref{ell12}, \ref{m12}).  These are the analogs of the `bubble' equations for solitons~\cite{Bena:2005va}.

Using the constraints on the parameters (\ref{constr1}, \ref{constr2}), it may now be verified that near the centres,
\be
\omega = \mathcal{O}(X_i^2) \td \phi_i^+ +\mathcal{O}(Y_i^2) \td \phi_i^-  \; ,
\ee
with higher order terms analytic in $X_i^2, Y_i^2$. Hence the 1-form $\omega$ is also smooth at the centres.   We have thus derived necessary and sufficient conditions for the spacetime metric to be smooth and timelike at the centres.

Next, we examine the Maxwell field \eqref{max} near the centres. From the discussion above $\td (f(\td t + \omega))$ is smooth at each centre. One can also verify that
\begin{eqnarray}\td \left[
-\frac{K}{H}(\td \psi + \chi) - \xi \right]
= \td [\mathcal{O}(X_i^2) \td \phi_i^+ +\mathcal{O}(Y_i^2) \td \phi_i^- ] \; ,
\end{eqnarray}  
with subleading terms analytic in $X_i^2, Y_i^2$; hence this 2-form is smooth at the centres. We deduce that the full Maxwell field is smooth at the centres.

Next we consider regularity of the spacetime away from the centres. It is clear from (\ref{f}) that $f$ is smooth everywhere away from the centres provided,
\be
K^2+HL>0  \; . \label{regular}
\ee
 Notice that this also implies that $f$ has the same zeroes as $H$. Remarkably, it turns out that (\ref{regular}) also guarantees that the full spacetime metric away from the centres is smooth and Lorentzian with a smooth inverse, even on surfaces where $f=0$. Furthermore, this condition also guarantees the Maxwell field is smooth away from the centres.  It is clear that (\ref{regular}) is satisfied asymptotically $r\to\infty$ and we will examine this condition in the interior later.

Let us now turn to causality of our spacetime.
We will require our spacetime to be stably causal with respect to the time function $t$. The condition for this is
 \bea
 g^{tt} &=& - f^{-2}+ fH \omega_\psi^2 + f H^{-1} \hat{\omega}_i \hat{\omega}_i  <0   \label{gtt} \; .
 \eea  
Clearly this is satisfied in the asymptotically flat region since $g^{tt} \to -1$ as $r\to \infty$. We will examine this condition in the interior explicitly later. At this stage we remark that generically it imposes inequalities on the parameters of the solutions.

To summarise, we have found a family of asymptotically flat spacetimes which are smooth even at the centres ${\bf x}_1$ and ${\bf x}_2$ and look like the origin of $\mathbb{R}^{1,4}$ near these points. The solutions are parameterised by $(a_1, a_2, k_0, k_1, k_2, \ell_0)$ subject to the constraints (\ref{constr1}, \ref{constr2}) and any inequalities arising from (\ref{regular}, \ref{gtt}). Thus, generically we have a 4-parameter family of solutions. Generically, the constraints can be easily solved for $(k_0, \ell_0)$ since they are linear in these parameters. However, we will refrain from doing so at this stage, since this involves imposing certain restrictions on the parameters. Later we will discuss various non-trivial special cases in which they can be solved.
 
 \subsection{Black hole and near-horizon geometry}
 \label{sec:horizon}
 
 We will now show that one can choose the parameters in our solution such that $r=0$ is an event horizon. As we will see, this actually imposes less constraints on the parameters, than had we imposed this to be a smooth centre. It is convenient to define the constants
 \be
j= k_0^2 +\ell_0, \qquad  \lambda = j - \frac{k_0^2(k_0^2 + \frac{3}{2} \ell_0)^2}{j^2}  \; .
\ee
Near $r \to 0$, to leading order we find
\be
K^2+HL= \frac{j}{r^2} + \mathcal{O}(r^{-1}), \qquad g^{tt} =  - \frac{\lambda j}{r^2} +O(r^{-1}) \; .
\ee 
Hence our spacetime is smooth and stably causal near $r=0$ if and only if $j >0, \lambda>0$. Furthermore, we will show that $r=0$ is a regular horizon if and only if
\be
j>0, \qquad \lambda>0 \; ,   \label{regularhorizon}
\ee
are both satisfied.

To this end, we transform to new coordinates $(v,r, \psi', \theta, \phi)$ defined by
\be
 \td t = \td v +\left( \frac{A_0}{r^2} + \frac{A_1}{r} \right) \td r  \qquad \td\psi = \td\psi' +\frac{B_0}{r} \td r
 \ee
where $A_0,A_1,B_0$ are all constants.  We will need the following Laurent expansions about $r=0$,
 \bea
 && f^2 = \frac{r^2}{j^2} + \frac{ \beta r^3}{j^3} +\mathcal{O}(r^4)\\ 
 &&\frac{H}{f} =  \frac{j}{r^2}  + \frac{\gamma}{r} + \mathcal{O}(1) \; , \qquad (Hf)^{-1} = j + \tilde{\gamma} r+ \mathcal{O}(r^2) \\
 && \chi_\phi = \cos \theta +\mathcal{O}(r) \; , \qquad \omega_\psi  =   \frac{k_0(k_0^2 + \frac{3}{2}\ell_0)}{r}  + \delta + \mathcal{O}(r) \; , \qquad \hat{\omega}_\phi = \mathcal{O}(r)
\eea
where $\beta, \gamma, \tilde{\gamma}, \delta$ are {\it constants} whose precise form we will not need.  It is then easy to check
\bea
&&g_{vv} = - f^2 = - \frac{r^2}{j^2} +\mathcal{O}(r^3) \; , \qquad  g_{v\psi'} = -f^2 \omega_\psi=  -\frac{r k_0(k_0^2 + \frac{3}{2}\ell_0)}{j^2} + \mathcal{O}(r^2) \; , \\
&&g_{\psi' \psi'} = \lambda + \mathcal{O}(r)   \; .
\eea 
The $g_{rr}$ component of the metric contains singular terms $1/r^2$ and $1/r$, whereas the $g_{r \psi'}$ component contains $1/r$ terms. Demanding that the $1/r^2$ term in $g_{rr}$ and the $1/r$ term in $g_{r \psi'}$ vanish, is equivalent to choosing the constants $A_0, B_0$ to be 
\be
A_0^2 = j^2 \lambda, \qquad B_0 = \frac{k_0 (k_0^2 + \frac{3}{2} \ell_0)}{j^2 \lambda} A_0  \; .
\ee
This gives
\be
g_{vr} = \pm \frac{1}{\sqrt{\lambda}} + \mathcal{O}(r)  \; , \qquad g_{\psi' r} = \mathcal{O}(1) \; ,
\ee
where the sign of $g_{vr}$ is positive (negative) if $A_0<0$ ($A_0>0$).
Finally, demanding that the $1/r$ term in $g_{rr}$ is absent determines $A_1$ to be
\be
A_1  = \frac{j\lambda}{2 A_0} \left( \gamma+ \tilde{\gamma} B_0^2 - \frac{\beta j}{\lambda} \right)  - B_0 \delta \; .
\ee
We now have 
\be
g_{rr} = \mathcal{O}(1)  \; .
\ee
The metric and its inverse are now analytic at $r=0$ and can therefore be extended to a new region $r<0$. The supersymmetric Killing field $V = \partial / \partial v$ is null  at $r=0$ and
\be
V_\mu \td x^\mu |_{r=0} = \pm \left( \frac{\td r}{\sqrt{\lambda}} \right)_{r=0}  \; .
\ee
This shows that the hypersurface $r=0$ is a degenerate Killing horizon of $V$. The upper sign choice corresponds to a future horizon, whereas the lower sign choice to a past horizon.

It remains to verify that the Maxwell field is regular on the horizon. In fact the gauge field $A$ may be read off from (\ref{max}). Changing to the above coordinates we find
 \be
 A_v = f = \frac{r}{j} +\mathcal{O}(r) \; , \qquad  A_{\psi'} = \frac{k_0(k_0^2+\tfrac{3}{2} \ell_0)}{j}-k_0 + \mathcal{O}(r)\; .
 \ee
 The $A_r$ component of the gauge field has a singular $1/r$ term; however the coefficient multiplying this is a constant and therefore it is pure gauge. We conclude that there is a gauge in which $A$ is analytic at $r=0$ and therefore $F$ is analytic at $r=0$, as required.

The near-horizon geometry is now easily extracted from the above by replacing $(v,r)\to (v/ \epsilon, \epsilon r)$ and letting $\epsilon \to 0$. The result can be written as
\bea
 && \td s^2_{\text{NH
}} = -\frac{r^2 \td v^2}{j \lambda} \pm  \frac{2 \td v \td r}{\sqrt{\lambda}}  + \lambda \left[ \td \psi' + \cos \theta \td \phi -\frac{k_0(k_0^2 + \frac{3}{2}\ell_0)}{j^2 \lambda}  r\td v \right]^2+ j (\td \theta^2+\sin^2\theta \td \phi^2)  \; ,  \nonumber \\
 && F_{\text{NH}}  =\frac{\sqrt{3}}{2} \td \left[ \frac{ r \td v}{j} +  \frac{k_0(k_0^2+\tfrac{3}{2} \ell_0)}{j} (\td \psi'+ \cos \theta \td \phi) \right] \; .
 \eea
 It is easy to see this is isometric to the near-horizon geometry of the BMPV black hole.  This was in fact guaranteed by a near-horizon uniqueness theorem proved in~\cite{Reall:2002bh}.
 
\subsection{Geometry of the axes}\label{subsec:axes}

The axes of rotation corresponds to the $z$-axis of the $\mathbb{R}^3$ base of the Gibbons-Hawking base. Due to the sources in the various harmonic functions, this axis naturally splits into four intervals: 
\bea
I_+ &=&\{ z\geq a_2 \},  \label{I+}\\    
I_C &=& \{ a_1\leq z \leq a_2 \},  \label{IC}\\ 
I_D &=& \{ 0\leq z \leq a_1 \} , \label{ID} \\
I_-&=& \{ z \leq 0 \}. \label{I-}
\eea
In polar coordinates (\ref{I+}, \ref{IC}, \ref{ID}) correspond to $\theta=0$, whereas (\ref{I-}) to $\theta=\pi$.   

It is easy to see that along this axis
\be
\chi = \pm \td \phi
 \ee
 where the $+$ sign occurs for $I_+$ and $I_D$ and the $-$ sign for $I_C$ and $I_-$. 
Remarkably, it can be checked that anywhere along the $z$-axis the one-form
\be
\hat{\omega}=0  \; .
\ee
This is a non-trivial identity which requires use of the constraints (\ref{constr1}, \ref{constr2}) in various combinations depending on the interval one is in.   It then follows that the induced metric on the axis is
 \be
\td s^2_3 = -f^2( \td t +\omega_\psi \td\psi^\pm )^2  + \frac{H \td z^2}{f} + \frac{1}{fH} (\td \psi^\pm)^2   \label{axis}
\ee
where $\psi^\pm = \psi \pm \phi$ and $\phi^\pm=\phi$ depending on which interval we are in.  We also find the induced Maxwell field is
\be 
F_{3} = \frac{\sqrt{3}}{2} \td[  f \td t+ (f\omega_\psi  -  K H^{-1}) \td\psi^\pm]   \; .  \label{maxis}
\ee
Note that $\partial_{\psi^\pm}= \partial_\psi$ and $\partial_{\phi^\pm} = \partial_\phi \mp \partial_\psi$ and
\be \label{positivity}
|\partial_{\psi^\pm} |^2 = \frac{1}{fH} - f^2 \omega_\psi^2  \; .
\ee
We remark that the stably causal condition (\ref{gtt}) on the axis reduces to just $-f^{-2} + fH \omega_\psi^2<0$. Therefore, we deduce that, at least away from fixed points of $\partial_{\psi^\pm}$, stable causality follows from simply imposing $| \partial_{\psi^\pm} |>0$.

We will now examine the geometry on the axis in detail. The precise details depend on the interval in question, although in all cases we show that regularity and causality reduce to the positivity of various polynomials in the given interval. Later we will examine various special cases where one can demonstrate positivity of the various above polynomials is equivalent to certain inequalities on the parameters.

\subsubsection{Semi-infinite axes}
The semi-infinite intervals $I_\pm$ correspond to the two axes of rotation which extend out to asymptotic infinity. In particular, along $I_+$ the Killing field $\partial_{\phi^+}=0$, whereas along $I_-$ it is $\partial_{\phi^-}=0$. At the endpoint $z=a_2$ of $I_+$ the Killing field $\partial_{\psi^+}$ must degenerate smoothly due to the general regularity conditions we imposed at the centres. On the other hand, at the endpoint $z=0$ of $I_-$ this axis meets the horizon.

We now examine the geometry on $I_+$ in detail. At $z=a_2$ the Killing field $\partial_{\psi^+}$ degenerates resulting in a conical singularity; it is easy to see this is removable since $\Delta \psi^+=4\pi$. Now consider $z > a_2$.  One finds, using (\ref{ell12}),
\begin{equation}
H = \frac{z^2  -2 a_1 z + a_1 a_2}{z(z-a_1)(z-a_2)}\,, \qquad f = \frac{z(z^2-2a_1 z + a_1 a_2)}{P_+(z)}\; ,
\end{equation} where $P_+(z)$ is a cubic. Observe that $H>0$ for $z>a_2$. Smoothness of $f$ requires $P_+(z)>0$ and therefore we also deduce $f>0$  on $I_+$.  We also find,
\begin{equation}
|\partial_{\psi^+}|^2 = \frac{Q_+(z)}{P_+(z)^2}  \; ,
\end{equation} where $Q_+(z)$ is a 7th-order polynomial. Therefore we also require $Q_+(z) >0$ for all $z > a_2$.    We thus deduce that (\ref{axis}) is a smooth Lorentzian metric for $z > a_2$ if and only if $P_+(z)>0, Q_+(z)>0$.   Finally, the Maxwell field (\ref{maxis}) is also smooth on $I_+$.  To see this, note $f$ is smooth and the following holds:
\be\label{Maxwellsmooth}
f \omega_\psi - K H^{-1} = \frac{R_+(z)}{P_+(z)} \; ,
\ee
where $R_+(z)$ is a cubic. Hence, given $P_+(z)>0$, this function is smooth for all $z >a_2$ and smoothness of the Maxwell field follows. 

Similar conclusions hold for $I_-$.  Consider first $z' = -z > 0$. One finds, using \eqref{ell12},
\begin{equation}
H = \frac{z'^2 + 2a_1 z' + a_1a_2}{z' (z'+a_2)(z' + a_1)} \, ,  \qquad f = \frac{z'(z'^2 + 2a_1 z' + a_1 a_2)}{P_-(z')} \; ,
\end{equation} where $P_-(z')$ is a cubic.  For $z'>0$, note that $H> 0$.  Smoothness of $f$ requires $P_-(z')>0$ and thus $f>0$ for $z'>0$. Also we find,
\begin{equation}
|\partial_{\psi^{-}}|^2 = \frac{Q_-(z')}{P_-(z')^2} \; ,
\end{equation} where $Q_-(z')$ is a 7th-order polynomial. Thus we also require $ Q_-(z') > 0$  for $z'>0$. We deduce that (\ref{axis}) is a smooth Lorentzian metric for $z <0$ if and only if $P_-(z')>0, Q_-(z')>0$.  As $z' \to 0$ we see $f$ vanishes; this corresponds to where $I_-$ meets the event horizon.  Turning to the Maxwell field, one finds that the identity \eqref{Maxwellsmooth} holds with $P_+(z)$ and $R_+(z)$ replaced by $P_-(z')$ and a cubic $R_-(z')$ respectively.  Therefore, the requirement $P_-(z')>0$ is sufficient to ensure smoothness of the Maxwell field on $I_-$. 

\subsubsection{Bubble}
Now consider the interval $I_C$. Along this interval $\partial_{\phi^-}=0$, whereas $\partial_{\psi^-}$ must degenerate at the endpoints $z=a_1, a_2$ smoothly due to regularity at these centres. To see this explicitly define $\rho^2= |z-a_i|$ and $f_i = f_{z=a_i}$ for $i=1,2$. We then find that as $\rho \to 0$
\be
\td s_3^3 \sim -f_i^2 \td t^2 + O(\rho^2) \td t \td \psi^-  + \frac{4}{|f_i|} \left[  \td \rho^2 +\frac{\rho^2}{4} (\td \psi^-)^2 \right]
\ee
is indeed free of conical singularities at both endpoints since from above $\Delta \psi^- = 4\pi$.  However, since regularity at the centres requires $f_{1}<0$ and $f_{2}>0$ there must be a point $z_0 \in (a_1,a_2)$ where $f_{z=z_0}=0$.  Using (\ref{ell12}) we find that
\be
f = \frac{z (z^2 -a_1 a_2)}{P_C(z)}
\ee
where 
\bea
\label{PCgen}
P_C(z) &=& z^3 + [\ell_0-k_0^2+2(k_1k_2+k_0k_2-k_0k_1)]z^2 \nonumber \\ &+&[a_1(k_0-k_2)^2+ a_2(k_0+k_1)^2-a_1a_2]z -a_1 a_2 (k_0^2+\ell_0)  \; .
\eea
 Therefore, we deduce that $z_0= \sqrt{a_1 a_2}$ and $P_C(a_i)>0$ to ensure the correct signs of $f$ at the endpoints.  Furthermore, smoothness of the invariant $V^2=-f^2$ requires $P_C(z)>0$ for all $z\in [a_1, a_2]$. We also find,
\be
H = \frac{z^2-a_1a_2}{z(z-a_1)(a_2-z)} \; ,
\ee
which therefore also changes sign in the interval in such a way that $g_{zz}=H/f>0$ and smooth for $z\in (a_1, a_2)$. Next, we find the remarkable simplification
\be
|\partial_{\psi^-} |^2 = \frac{Q_C(z)}{P_C(z)^2} \; ,
\ee
where $Q_C(z)$ is a complicated 7th order polynomial. Therefore smoothness requires $Q_C(z)>0$ for all $z \in (a_1, a_2)$.  Putting all this together, we find that (\ref{axis}) is a smooth Lorentzian metric for all $z\in (a_1, a_2)$ if and only if $P_C(z)>0$ and $Q_C(z)>0$.  

Given these inequalities, we have shown that any constant time slice extends to a smooth inhomogeneous $S^2$. Therefore our black hole solution possesses a non-trivial 2-cycle $C$ in the domain of outer communication. 

Furthermore, the Maxwell field (\ref{maxis}) is also smooth on $C$. To see this it is sufficient to note the identity
\be
f \omega_\psi - K H^{-1} = \frac{R_C(z)}{P_C(z)} \; ,
\ee
where $R_C(z)$ is a cubic. Therefore, given $P_C(z)>0$, this function is smooth for all $z \in [a_1,a_2]$, which guarantees the Maxwell field is smooth on $C$.

\subsubsection{Disc}
Finally consider the interval $I_D$.  Along this interval $\partial_{\phi^-}=0$, whereas $\partial_{\psi^-}$ must degenerate at the endpoint $z=a_1$ smoothly due to regularity at this centre. Indeed defining $\rho^2= |z-a_1|$ we find that as $\rho \to 0$
\be
\td s_3^3 \sim -f_1^2 \td t^2 + O(\rho^2) \td t \td \psi^+  + \frac{4}{|f_1|} \left[  \td \rho^2 +\frac{\rho^2}{4} (\td \psi^+)^2 \right]
\ee
is indeed smooth since $\Delta \psi^+ = 4\pi$.  On the other hand, as we approach the other endpoint $z\to 0$ we find
\be
| \partial_{\psi^+} |^2  = \frac{(4 \ell_0+ 3k_0^2)\ell_0^2}{4(\ell_0+k_0^2)^2} + \mathcal{O}(z)
\ee
so $\partial_{\psi^+}$ does not have a fix point there. In fact this endpoint corresponds to the horizon and it is easily see that this agrees with the norm in the near-horizon geometry.   Now, since $f>0$ just outside the horizon and $f_1<0$, again we deduce there must be a point in the interval $z_0\in (0,a_1)$ where $f=0$. In this case we find
\be
f =  \frac{z (z^2-2a_2 z+a_1a_2)}{P_D(z)}  \; ,
\ee
where $P_D(z)$ is a cubic. The quadratic $z^2-2a_2 z+a_1a_2$ is negative at $z=a_1$ and positive at $z=0$, with one root in between at $z_0  = a_2- \sqrt{a_2(a_2-a_1)}$. Hence we also require $P_D(z)>0$ for all $z\in (0, a_1]$. Similarly, we also find
\be
 H = \frac{z^2-2a_2 z +a_1a_2}{z(a_1-z)(a_2-z)} \; ,
 \ee
 and remarkably 
 \be
| \partial_{\psi^+} |^2 = \frac{Q_D(z)}{P_D(z)^2}
\ee
where $Q_D(z)$ is a 7th order polynomial. Thus we require $Q_D(z)>0$ for all $z \in (0,a_1)$. Putting all this together we find that again (\ref{axis}) is a smooth Lorentzian metric for all $z \in (0,a_1)$ if and only if $P_D(z)>0$ and $Q_D(z)>0$. 

Given these inequalities, we deduce that constant time slices extend to a smooth positive definite metric on a topologically disc surface $D$ in the spacetime, whose centre touches $C$ at a point ($z=a_1$) and ends on the horizon. However, note that since $H/f \sim j/z^2$ as $z\to 0$,  the proper radius of this disc is infinite; this of course should be the case since our horizon is degenerate.  

Again, it is easily checked the Maxwell field is smooth on $D$. Just note the identity
\be
f \omega_\psi - K H^{-1} = \frac{R_D(z)}{P_D(z)} \; ,
\ee
where $R_D(z)$ is a cubic. It follows that, given $P_D(z)>0$, this function is smooth for all $z \in (0,a_1]$, which guarantees the Maxwell field is also smooth on $D$. \\

We observe that from the above analysis we see that the supersymmetric Killing field $V$ is in fact null on the circle $z= \sqrt{a_1a_2}$ in $C$, and also on the circle $z=  a_2- \sqrt{a_2(a_2-a_1)}$ in $D$. This is why our black hole solution evades the black hole uniqueness theorem for BMPV~\cite{Reall:2002bh}, since that {\it assumes} $V$ is strictly timelike in the exterior region.

 \subsection{Physical properties}
 
 The conserved global charges of our asymptotically flat solution are easily computed. We find the electric charge is,
 \bea
 Q = \frac{1}{4\pi} \int_{S^3} \star F =2\sqrt{3} \pi [(k_0 + k_1 + k_2)^2 + \ell_0 + \ell_1 + \ell_2 ]  \; .
 \eea
 It is easily checked the mass $M= -\frac{3}{32\pi} \int_{S^3} \star d V =\frac{\sqrt{3}}{2} Q$ satisfies the BPS bound.  The angular momentum with respect to a rotational Killing field $m$ is given by $J[m] = \frac{1}{16\pi} \int_{S^3} \star dm$. We find that with respect to $\partial_\psi$ and $\partial_\phi$ it is given by,
 \bea
 &&J_\psi= 2\pi \left[(k_0+k_1+k_2)^3 +m_1+m_2 + \frac{3}{2}(\ell_0+\ell_1+\ell_2)(k_0+k_1+k_2) \right]  \; ,\\
 &&J_\phi = \pi \left[ 6 a_1 k_1 + 3 \left(a_1(k_2 + k_0) - a_2(k_0 + k_1)\right) \right]    \; ,\label{Jphi}
 \eea
 respectively.
 
It is easily seen that for any BPS black hole in minimal supergravity the surface gravity $\kappa$ and angular velocities $\Omega_i$  of the horizon all vanish, whereas the electric potential on the horizon $\Phi_H = \frac{\sqrt{3}}{2}$.  Also we find that the area of the horizon is
\be
A_H = 16\pi^2 j \sqrt{\lambda}
\ee
and the electric charge and angular momentum of the horizon are
\be
Q_H = 2 \sqrt{3} \pi \lambda
\ee
and
\be
J_\psi^H = -\frac{\pi \lambda k_0 (k_0^2+ \tfrac{3}{2} \ell_0)}{j}  , \qquad \qquad J_\phi^H=0 \; ,  \label{Jhorizon}
\ee
respectively. Thus we see that the parameters $k_0, \ell_0$ are related to the conserved charges on the horizon.
 
Our black hole solution also carries local magnetic flux charge due to the presence of a non-trivial 2-cycle $C$ in the domain of outer communication of the black hole. Using the expression for the Maxwell field induced on this bubble (\ref{maxis}) we find the magnetic flux is
\be
q[C] = \frac{1}{4\pi} \int_{C} F  = -\frac{\sqrt{3}}{2} (k_1+k_2)  \; .
\ee
Furthermore, due to the presence of a horizon and 2-cycle, a disc topology region $D$ connecting the two also exists on which we can define a flux charge~\cite{Kunduri:2013vka}. Similarly, we find this is
\be
q[D] = \frac{1}{4\pi} \int_{D} F = \frac{\sqrt{3}}{2} \left(k_1 +k_0 - \frac{k_0(k_0^2+\tfrac{3}{2} \ell_0)}{j} \right)  \; .
\ee
This gives a physical interpretation to the parameters $k_1, k_2$ in our solution. 

In fact, one can show the flux $q[C]$ is necessarily non-zero. To see this, set $q[C]=0$ so $k_2=-k_1$. Then the constraints (\ref{constr1}, \ref{constr2}) imply that $k_1=-k_0$, and either $k_1=0$ or $\ell_0= - \tfrac{2}{3} k_0^2$. In either of these cases, the polynomial (\ref{PCgen}) has a factor of $z^2-a_1a_2$, so that it vanishes at $z=\sqrt{a_1 a_2}$, which means the solution is not smooth on the 2-cycle $C$.  
This is consistent with the physical expectation that the flux `supports' the 2-cycle $C$ from collapsing.

Recently, a Smarr relation and first law of black hole mechanics were derived for black hole spacetimes with non-trivial topology, see equations (\ref{BHmass}, \ref{BHmech}). For minimal supergravity, the `electric' flux charge appearing in these laws is given by
\be
\mathcal{Q}[C] = \frac{1}{2} \int_C \left( i_V \star F + \frac{8}{\sqrt{3}} (\Phi_H - \Phi) F \right) \; ,
\ee
and similarly for $\mathcal{Q}[D]$. Using the general form for a supersymmetric solution~\cite{Gauntlett:2002nw}, it is a simple exercise to check the integrand $i_V \star F + \frac{8}{\sqrt{3}}  (\Phi_H - \Phi) F = \sqrt{3} \td ( f^2 (\td t+\omega))$. Then, using the behaviour of the solution near the various centres, it follows that
\be
\mathcal{Q}[C]  =0 \; , \qquad \qquad  \mathcal{Q}[D] =0  \; .
\ee
Therefore, the extra terms in the Smarr relation and first law in fact vanish for such supersymmetric black holes. Thus these mass formulae reduce to the standard BPS bound.

 \subsection{Static horizon}
 \label{sec:static}
 
Consider the special case in which the horizon carries no angular momentum. From (\ref{Jhorizon}), we deduce that this special case can be achieved by setting\footnote{Another possibility is $\ell_0 = - \tfrac{2}{3} k_0^2$, which also has a regular horizon. We have not investigated this case.}
\be
k_0=0  \; .
\ee
Inspecting the near-horizon geometry of the general solution, we see that in this case it reduces to the static AdS$_2\times S^3$ near-horizon geometry, as one would expect. Thus we obtain a three parameter family of stationary, but non-static, black hole solutions with the same near-horizon geometry as extreme Reissner-Nordstr\"om solution. This illustrates the point that a static near-horizon geometry~\cite{Kunduri:2009ud} need not arise as the near-horizon limit of a static black hole. We believe this is the first such example with spherical horizon topology.\footnote{The BPS black ring is non-static, but possesses a static  locally AdS$_3\times S^2$ near-horizon geometry~\cite{Elvang:2004rt}.}

In fact, further setting, 
\be
k_2=0 \; ,
\ee 
gives a two parameter family which is particularly amenable to analysis. Observe that physically, this corresponds to setting the magnetic flux through the bubble and disc to be equal and opposite, $q[C]=-q[D]$.  Indeed, in this case we may solve the constraints (\ref{constr1}, \ref{constr2}) explicitly to find
\bea
&&k_1^2= 3(a_2-a_1) \; , \\
&&\ell_0 = a_2 -2a_1  \; .
\eea
We see that the conditions for regularity of the horizon (\ref{regularhorizon}) reduce to simply $\ell_0>0$. Therefore, we deduce a stronger inequality on the parameters,
\be
a_2 > 2a_1  \; .  \label{ineq}
\ee 
For definiteness we will take the positive root $k_1>0$.

We now turn to the regularity conditions for the axis, derived in Section \ref{subsec:axes}. In particular, we have proved that in this special case, the polynomials $P_\pm(z), Q_\pm(z), P_C(z), Q_C(z), P_D(z)$ and  $Q_D(z)$ are strictly positive in their respective intervals, provided (\ref{ineq}) is satisfied.  This establishes smoothness of the spacetime on the bubble $C$ and the disc $D$ and the two semi-infinite axes $I_\pm$. As discussed in Section \ref{subsec:axes}, these conditions are also sufficient to ensure stable causality on the whole $z$-axis (i.e. $\theta \in \{0,\pi\}$). 

We have not been able to prove smoothness and stable causality away from the axis for general values of $r$ (of course in the two asymptotic regions $r\to \infty$ and $r \to 0$ we have showed this earlier for the general solution).  However, we have performed extensive numerical checks and not found any violation of the inequalities (\ref{regular}) and $g^{tt}<0$ for this solution. In brief, our strategy was to treat $K^2+HL$ and $g^{tt}$ as a function of four variables $(r,\theta, a_1, a_2)$ and sample it over a variety of hyper-grids in the domain $r>0$, $0<\theta<\pi$ and $a_2>2a_1>0$.  Therefore, we believe that this solution is indeed smooth and stably causal if and only if (\ref{ineq}) is satisfied.

It is worth noting some of the physical properties of this special case. Although the near-horizon geometry is not rotating, the black hole spacetime in fact possesses non-zero angular momenta given by
\begin{equation}
J_\psi = 3k_1 \pi(7a_2 - 8a_1) \; ,\qquad J_\phi = -3k_1 \pi (a_2 - 2a_1) \; , 
\end{equation} 
from which it follows that $J_\psi>0$ and $J_\phi<0$.  As we discuss in the next section, this in particular means that this black hole spacetime never possesses equal angular momenta with respect to the orthogonal $U(1)^2$ Killing fields of $\mathbb{R}^4$.  Therefore, its conserved charges are never the same as that of the BMPV black hole (including the non-rotating Reissner-Nordstr\"om solution).

 \subsection{Equal angular momenta and black hole non-uniqueness}
 \label{sec:equalJ}
 
Here we consider solutions with equal angular momenta with respect to the standard $U(1)^2$ rotational Killing fields of $\mathbb{R}^{1,4}$. Recall that the BMPV black hole solution has this property~\cite{BMPV, Gauntlett:1998fz}. Define the quantity
\be
\eta =  \frac{Q^3}{ 24 \sqrt{3} \pi }  -\frac{J_\psi^2}{4}  \; .
\ee
The BMPV solution exists only if it is `under-rotating' $\eta>0$. Interestingly, we will show that there exist regular black hole solutions with equal angular momenta which are under-rotating $\eta>0$, critical $\eta=0$, or over-rotating $\eta<0$.

In our coordinates equal angular momenta can be achieved by setting $J_\phi=0$. From (\ref{Jphi}) this allows us to solve for $k_0$:
\begin{equation}\label{Jphivanishing}
k_0 = \frac{(2a_1 - a_2)k_1 + a_1 k_2}{a_2 - a_1} \; .
\end{equation} 
We thus obtain a three-parameter family of black hole solutions with equal angular momenta and the same near-horizon geometry as the BMPV black hole. 

For simplicity consider the two-parameter subset of solutions with 
\be
k_2 =0 \; .
\ee  
In this case we must have $k_1 \neq 0$ as otherwise the solution is singular. 
Then the unique solution to the constraints is given by
\bea
&& k_1^2 = 3a_1 \\
&&\ell_0 = \frac{3a_1(2a_1 - a_2)}{a_2 - a_1}  \; .
\eea 
The conditions for regularity of the horizon (\ref{regularhorizon}) are positivity of the quantities
\begin{equation}
k_0^2 + \ell_0 = \frac{3a_1^2(2a_1 - a_2)}{(a_2-a_1)^2} \; , \qquad \qquad k_0^2  + \frac{4}{3} \ell_0 = \frac{a_1(2a_1 - a_2)(2a_1 + a_2)}{(a_2 - a_1)^2} \; .
\end{equation}
Therefore we deduce these are met if and only if,
\be
a_1< a_2<2a_1 \;.  \label{domain}
\ee 
For definiteness we take the positive root $k_1>0$ (the choice $k_1 < 0$ leads to a solution with equal and opposite angular momenta).

The analysis of this solution is more involved than the special case described in the previous section.  We can prove that the polynomial $P_C(z)>0$, but for the remaining polynomials $Q_\pm, P_\pm , Q_C, P_D, Q_D$ we have relied on numerical methods to demonstrate positivity.  Turning to checking smoothness and stable causality away from the axis for general $(r,\theta)$, we have done a careful search for violations of (\ref{regular}) and $g^{tt} < 0$.  As in the previous case, the method is to sample a large number of points in the parameter space $(r,\theta,a_1,a_2)$ with $r>0, 0 < \theta < \pi$ and $0 < a_1 < a_2  < 2a_1$.  We have found no evidence for violations of  (\ref{regular}) or $g^{tt}<0$ and believe the solution is smooth and stably causal if and only if \eqref{domain} holds. 

Let us now briefly discuss the physical properties of this solution.  We find the angular momentum,
\begin{equation}
J_\psi = \frac{3\sqrt{3a_1^3}\pi}{(a_2 - a_1)^3}\left[a_2^3  - 3 a_1 a_2^2 + 6a_1^2 a_2 - 2a_1^3\right]  \; ,
\end{equation} and electric charge,
\begin{equation}
Q = \frac{6\sqrt{3}a_2 a_1^2\pi}{(a_2 - a_1)^2} \; .
\end{equation}
It is easily checked that in our parameter domain (\ref{domain}) $J_\psi>0$, so these solutions always have non-zero angular momentum.

Furthermore, it may be verified that
\bea
\eta=\frac{27 \pi^2 a_1^3\Big[ \sqrt{3} a_2 (2a_1-a_2) + 2(a_2-a_1)^2 \Big] \Big[ \sqrt{3} a_2 (2a_1-a_2) - 2(a_2-a_1)^2 \Big]}{4(a_2-a_1)^4}  \, .  \label{phase}
\eea
The first factor in the square brackets is always positive in our domain. Therefore the sign of (\ref{phase}) is the same as the sign of the quadratic in the second factor in square brackets. We thus find that (\ref{phase}) is positive if and only if
\be
 0.32\approx  1- \sqrt{-3+ 2\sqrt{3}} < \frac{a_2}{a_1} < 1+ \sqrt{-3 +2\sqrt{3}}  \approx 1.68  \; .  \label{critical}
\ee
Remarkably, this overlaps with our parameter domain (\ref{domain}). Therefore, we have shown there exist black hole solutions with the same conserved charges as the BMPV black hole, the same near-horizon geometry, but are distinct to the BMPV black hole.  

This explicitly demonstrates  non-uniqueness of supersymmetric black holes in minimal supergravity. Furthermore, in the complement of (\ref{critical}) within the domain (\ref{domain}) we may also have black holes for which (\ref{phase}) vanishes or is negative. Therefore we have also shown there exist spherical black hole solutions which fill out parts of the $(Q,J_\psi)$-phase space not occupied by the BMPV solution.

\subsection{A decoupling limit}

Above we showed that the near-horizon geometry of our black hole solution is the same as that of the BMPV black hole. In particular, the non-trivial 2-cycles in the exterior to the black hole are not captured in this limit. Indeed, a general feature of the near-horizon limit of extremal black holes is that it only retains properties of the spacetime intrinsic to the horizon~\cite{Kunduri:2013gce}.  It is natural to wonder if there is a limit of the solution  which focuses near the horizon in such a way to keep the 2-cycles in the exterior. Indeed, it is easy to write down a decoupling limit with this property.\footnote{Decoupling limits which preserve the number of centres in the harmonic functions have been previously considered, see e.g. ~\cite{deBoer:2008fk}. We thank Joan Simon for pointing this out.}

Let $\epsilon>0$ and define new coordinates and parameters by
\be 
t = \frac{\bar{t}}{\epsilon}, \qquad r = \epsilon \bar{r}, \qquad a_i = \epsilon \bar{a}_i
\ee
leaving all other coordinates and parameters the same. It is then easy to see that the solution is equivalent to that constructed from the Gibbons-Hawking base
\be
\overline{\td s}^2_M = \bar{H}^{-1} (\td \psi + \bar{\chi}_\phi \td \phi)^2 + \bar{H} ( \td \bar{r}^2 + \bar{r}^2 \td \Omega_2^2 )\; , \label{GHbar}
\ee
and the four harmonic functions
\begin{eqnarray}
\bar{H} &=& \frac{1}{\bar{r}}  - \frac{1}{\bar{r}_1}+ \frac{1}{\bar{r}_2}, \qquad  \qquad \bar{K} = \frac{k_0}{\bar{r}} + \frac{k_1}{\bar{r}_1} + \frac{k_2}{\bar{r}_2} ,\\
\bar{L}_\epsilon &=& \epsilon+ \frac{\ell_0}{\bar{r}}  + \frac{\ell_1}{\bar{r}_1} + \frac{\ell_2}{\bar{r}_2}, \qquad  \qquad \bar{M}_\epsilon =\epsilon m + \frac{m_0}{\bar{r}}  + \frac{m_1}{\bar{r}_1}+ \frac{m_2}{\bar{r}_2},
\end{eqnarray} 
where we have defined $\bar{r}_i$ as in (\ref{centres}) with $r \to \bar{r}$ and $a_i \to \bar{a}_i$, provided the integration constants in $\bar{\chi}, \bar{\hat{\omega}}, \bar{\xi}$ are chosen to be the same as those in ${\chi}, {\hat{\omega}}, {\xi}$ under the replacement $a_i= \epsilon \bar{a}_i$.

We may now take the limit $\epsilon \to 0$. Observe that this leaves $\bar{H}$ and $\bar{K}$ unchanged, whereas
\be 
\bar{L}_\epsilon \to \bar{L}  = \frac{\ell_0}{\bar{r}}  + \frac{\ell_1}{\bar{r}_1} + \frac{\ell_2}{\bar{r}_2}, \qquad  \qquad \bar{M}_\epsilon \to \bar{M} = \frac{m_0}{\bar{r}}  + \frac{m_1}{\bar{r}_1}+ \frac{m_2}{\bar{r}_2}. 
\ee
The limiting solution is thus equivalent to that constructed from the GH base (\ref{GHbar}) and
the harmonic functions $\bar{H}, \bar{K}, \bar{L}, \bar{M}$, with the integration constants in $\bar{\chi}, \bar{\hat{\omega}}, \bar{\xi}$ chosen to be the $\epsilon \to 0$ limits of those for the full solution. In particular, the limit of the 1-form $\hat{\omega}$ is given by
\begin{eqnarray}
\bar{\hat{\omega}}_\phi &=& \frac{1}{2\bar{r} \bar{a}_1}\left[(m_1+\tfrac{3}{2}(  \ell_1 k_0 -\ell_0 k_1)) \left(\bar{r}_1+ \frac{(\bar{r}^2-\bar{a}_1^2)}{\bar{r}_1} \right)\right]  \nonumber \\
&+&\frac{1}{2 \bar{r} \bar{a}_2}\left[ (m_2+\tfrac{3}{2}( \ell_2 k_0 -\ell_0 k_2))\left(\bar{r}_2+\frac{(\bar{r}^2-\bar{a}_2^2)}{\bar{r}_2} \right) \right]     \nonumber \\
&-&\frac{(m_1 + m_2+\tfrac{3}{2}(\ell_1 k_2 - \ell_2 k_1))}{\bar{r}_1\bar{r}_2(\bar{a}_2 - \bar{a}_1)}\left[\bar{a}_1 \bar{a}_2 - (\bar{a}_1 + \bar{a}_2)\bar{r} \cos\theta +\bar{r}^2 \right]  + \bar{c} \; ,
\end{eqnarray} 
where $\bar{c}$ is the `barred' version of the constant (\ref{c2_AF}). It is clear this limiting solution belongs to the family of solutions one would obtain by simply dropping the constants `$1$' and `$m$' in the original harmonic functions $L,M$.

Let us now examine the $\bar{r} \to \infty$ asymptotics of this new solution. It is easy to show that
\be
\bar{f} = \frac{\bar{r}}{j'} + \mathcal{O}(1), \qquad  \bar{\omega}_\psi  = \frac{\delta'}{\bar{r}} + \mathcal{O}(\bar{r}^{-2}), \qquad \bar{\hat{\omega}}_\phi = \mathcal{O}(\bar{r}^{-2})  \; ,
\ee
where we have defined the constants
\bea
j ' &=& \ell_0 +\ell_1+\ell_2 + (k_0+k_1+k_2)^2 \; , \\
\delta' &=& \frac{3}{2}( \ell_0 +\ell_1+\ell_2)(k_0+k_1+k_2) +  (k_0+k_1+k_2)^3+ m_1+m_2  \; .
\eea
This allows us to deduce the $\bar{r} \to \infty$  asymptotics of our new spacetime to be
\bea
&& \td s^2 = \left( -\frac{\bar{r}^2}{j'^2} + \mathcal{O}(\bar{r}) \right) \td \bar{t}^2 +\left(  \frac{j' }{\bar{r}^2} +\mathcal{O}(\bar{r}^{-3})\right) \td \bar{r}^2 + \left( - \frac{2\delta' \bar{r}}{j'^2} +\mathcal{O}(1) \right)\sigma_3 \td \bar{t} + \mathcal{O}(1) \td \phi \td \bar{t} \nonumber \\ && \qquad + \left( \lambda' + \mathcal{O}(\bar{r}^{-1} )\right) \sigma_3^2 + \left(  j' + \mathcal{O}(\bar{r}^{-1} )\right)  \td \Omega_2^2  + \mathcal{O}(\bar{r}^{-1}) \sigma_3 \td \phi +  \mathcal{O}(\bar{r}^{-2}) \td \phi^2   \; ,
 \eea
 where $\sigma_3 = \td \psi+ \cos \theta \td \phi$ and we have defined the constant
 \be
 \lambda' =  j' - \frac{\delta'^2}{j'^2}  \; .
 \ee
We recognise the leading order metric as the near-horizon geometry for the BMPV black hole (written in Poincar\'e AdS$_2$ coordinates). Indeed, comparing to section \ref{sec:horizon}, we see that it a BMPV near-horizon geometry with the constants $j \to j'$ and $ k_0 (k_0^2+ \tfrac{3}{2} \ell_0) \to \delta'$. Notice that this is equivalent to the replacements
\be
k_0 \to k_0+k_1+k_2, \qquad \ell_0 \to \ell_0+\ell_1 +\ell_2 \; .  \label{flow}
\ee

Let us now turn to regularity of our spacetime. One can repeat all the steps performed for the black hole solution and one finds it is smooth with the same topology provided the parameters satisfy (\ref{ell12}), (\ref{m12}) and
\be
\label{constr1dec}
 \bar{a}_1(k_2 + k_1)^3 + (\bar{a}_2 - \bar{a}_1)(3k_0 k_1^2 +k_1^3   -3k_1 \ell_0) = 0 \; , 
 \ee
 and
 \be
 \label{constr2dec}
 \bar{a}_2(k_2 + k_1)^3 + (\bar{a}_2 - \bar{a}_1)(3k_0 k_2^2 - k_2^3 + 3k_2 \ell_0) = 0  \; ,
\ee
together with inequalities arising from the smoothness and stably causal conditions (\ref{regular}) and (\ref{gtt}). Hence we again have a 4-parameter family of solutions. 
In fact this is guaranteed by regularity of the black hole solution for $\epsilon >0$ and continuity in the limit $\epsilon \to 0$. Indeed, the constraints on the parameters (\ref{constr1dec}, \ref{constr2dec}) are limits of the constraints (\ref{constr1}, \ref{constr2}).  Similarly, our spacetime possesses a regular degenerate horizon with the same near-horizon geometry as obtained in section \ref{sec:horizon}. Again, this can be shown by repeating the near-horizon analysis, or more simply appealing to continuity of our limit.

It is worth noting that not all solutions to the regularity constraints (\ref{constr1}, \ref{constr2}) of the asymptotically flat solution admit the above decoupling limit. For example, the special cases we examined in sections (\ref{sec:static}) and (\ref{sec:equalJ}) do not, because scaling the $a_i$ to zero is not compatible with keeping the $k_i$ fixed for those solutions. It may be interesting to investigate other decoupling limits, which in particular are well defined for these special cases.

To summarise, we have obtained a spacetime with the following properties. It is asymptotically the near-horizon geometry of a BMPV solution, possesses an event horizon with a near-horizon geometry of a {\it different} BMPV solution (with parameters related by (\ref{flow})), and a 2-cycle in the exterior region. In effect we have thus merely `decoupled' the asymptotically flat region from our black hole solution, while keeping the exterior topology and near-horizon region intact.  We emphasise that this solution is different to the decoupling limit of a multi-centred BMPV  black hole in that it represents a \emph{single} black hole in a near-horizon BMPV background. We also note it differs from the  smooth {\it solitons} that are asymptotic to a near-horizon BMPV solution previously obtained~\cite{Bobev:2011kk}.  

Black holes in spacetimes asymptotic to near-horizon geometries of black holes have of course been previously obtained. For example, multi-black holes in AdS$_2\times S^2$\cite{Maldacena:1998uz} and AdS$_3 \times S^2$~\cite{deBoer:2008fk} have been constructed. More recently, a 2-centred solution describing a single black hole in AdS$_3 \times S^2$ has been written down~\cite{Bena:2011zw}. 
 \section{Discussion}

We have presented an asymptotically flat supersymmetric black hole solution to minimal supergravity with spherical horizon topology distinct from the well known BMPV black hole~\cite{BMPV}. In contrast to the BMPV black hole, the topology of the domain of outer communication is non-trivial and given by (\ref{Sigma}) with $n=1$ and $n'=0$.
 We have also showed there are regimes in parameter space where the two solutions have identical mass, electric charge and angular momentum. The solutions may be distinguished by local magnetic flux charges defined on the non-trivial 2-cycles in the spacetime.  This is the first proof of non-uniqueness of supersymmetric black holes with a connected horizon in minimal supergravity.\footnote{The BMPV black hole~\cite{BMPV} and BPS black ring~\cite{Elvang:2004rt} never possess overlapping conserved charges and are each uniquely specified by these charges. On the other hand, in $U(1)^3$-supergravity the BPS black rings also carry independent dipole charges which is sufficient to exhibit non-uniqueness~\cite{Elvang:2004ds}.} Furthermore, we believe this provides the first explicit counterexample to {\it spherical} black hole uniqueness (for connected horizons) in a theory containing only a Maxwell field.
 
In fact a uniqueness theorem for the supersymmetric black holes with spherical horizon topology has been demonstrated~\cite{Reall:2002bh}, which assumes the supersymmetric Killing field $V$ is strictly timelike everywhere outside the black hole. This is not the case for our solution; $V$ is null on a circle on the bubble exterior to the black hole, which is how it may evade this uniqueness theorem.  We observe that this is not so surprising, due to the fact that soliton spacetimes are also not strictly stationary but have regions where $V$ is null, referred to as `evanescent horizons'~\cite{Gibbons:2013tqa}.  

Although we have only focused on a simple example, it is clear there are more general possibilities. Working within minimal supergravity, it is a simple matter to add more smooth centres to the Gibbons-Hawking base, following the method for pure solitons~\cite{Bena:2005va}. This should result in large classes of black holes with more general 2-cycle structure in the exterior region. Furthermore, it should be possible to construct black rings with exterior bubbles using the same method.

More broadly, and as argued in section~\ref{sec:topology}, analogous solutions should exist for non-supersymmetric black holes both within minimal supergravity but also other five-dimensional Einstein-Maxwell type theories.  This could be of particular interest since non-trivial spacetime topology plays a role in black hole thermodynamics~\cite{Kunduri:2013vka}.  Clearly, however,  in the absence of supersymmetry, constructing explicit solutions will be significantly harder. Perhaps progress for extremal, but non-supersymmetric, black holes can be made along the lines of \cite{Bena:2009a, Bena:2009b}.

An interesting open question is whether vacuum spacetimes can `support' non-trivial topology. We emphasise there are no theorems ruling out this possibility. In the context of Weyl solutions this would correspond to taking more complicated spacelike rod structure than has been previously considered.

It would be interesting to investigate the implications of our results for the string theory derivation of the entropy of the BMPV black hole \cite{BMPV, David:2002wn}. Implicit in these calculations is the assumption that black holes are uniquely specified by their conserved charges. We have demonstrated explicitly this assumption is false, even for spherical black holes in minimal supergravity. 

We have also written down a decoupling limit of our solution which focuses near the horizon while retaining the non-trivial topology of the spacetime.  It is worth emphasising that this decoupling limit may be of interest in its own right.  This spacetime can be thought of as interpolating between the near-horizon geometries of  two {\it different} BMPV black holes. It would be interesting to interpret this as a holographic renormalisation group flow in the context of AdS$_2$/CFT~\cite{Strom:1999, Sen:2008}. 

Our decoupling limit  can also be thought of as a black hole sitting in the near-horizon geometry of a BMPV black hole.  It is natural to expect non-BPS generalisations of such solutions, presumably arising as the decoupling limit of a non-BPS version of our asymptotically flat black hole solutions. Such geometries may indicate that the non-existence results obtained for finite-energy excitations of AdS$_2\times S^2$ and the near-horizon geometry of extremal Kerr~\cite{Maldacena:1998uz, Dias:2009ex, Amsel:2009ev, Hajian:2014twa}, do not generalise to theories in which black hole uniqueness fails.  \\

\noindent  {\bf Acknowledgements} \\

\noindent HKK is supported by an NSERC Discovery Grant. JL is supported by an EPSRC Career Acceleration Fellowship. We would like to thank Harvey Reall and Joan Simon for useful comments. This research was supported in part by Perimeter Institute for Theoretical Physics. Research at Perimeter Institute is supported by the Government of Canada and by the Province of Ontario.

\end{document}